\documentstyle[preprint,aps,epsf]{revtex}

\begin{document}
\draft
\preprint{KANAZAWA 95-14,
\ October, 1995}

\title{Monopoles and Spatial String Tension
in the High Temperature Phase of SU(2) QCD}

\author{Shinji Ejiri
\footnote{ E-mail address : ejiri@hep.s.kanazawa-u.ac.jp} }

\address{Department of Physics, Kanazawa University, Kanazawa 920-11, Japan}

\maketitle

\begin{abstract}
We studied a behavior of monopole currents in the high temperature
(deconfinement) phase of abelian projected finite temperature SU(2) QCD
in maximally abelian gauge.
Wrapped monopole currents closed by periodic boundary play
an important role for the spatial string tension
which is a non-perturbative quantity in the deconfinement phase.
The wrapped monopole current density seems to be non-vanishing
in the continuum limit.
These results may be related to Polyakov's analysis of
the confinement mechanism using monopole gas in 3-dimensional SU(2)
gauge theory with Higgs fields.
\end{abstract}

\section{Introduction}

A characteristic feature of finite temperature QCD
is the deconfinement phase transition.
In the low temperature phase, the heavy quark potential
give rise to a confining (linear) potential
$ V(r) \sim \sigma r , $
where $\sigma$ is the string tension.
The string tension vanishes at the critical temperature
and the confining potential changes to a Debye screened potential
in the high temperature phase.
However, there are some non-perturbative quantities even in the high
temperature (deconfinement) phase of QCD.
The spatial string  tension is known to be such a non-perturbative quantity.
This quantity is the string tension extracted from
space-like Wilson loops composed of only spatial link variables.
The spatial string tension is non-vanishing even
in the high temperature phase.

This property in the high temperature phase may be understood by
dimensional reduction \cite{pisarski}.
4-dimensional QCD at high temperature limit can be regarded as
an effective 3-dimensional QCD with Higgs fields.
The effective theory has confinement features.
The string tension in this effective theory may be the spatial string
tension in 4-dimensional QCD.
The relation between the spatial string tension and that of the effective
theory is confirmed using Monte-Carlo simulations \cite{bali,karsch}.
The spatial string tension shows a scaling behavior on the lattice
and is non-vanishing in the continuum limit.
The behavior is reproduced by the following
obtained from the effective theory:
\begin{eqnarray}
\sqrt{\sigma_{s}} \propto g^{2}(T)T ,
\end{eqnarray}
where $ \sigma_{s} $ is the spatial string tension.
$ g(T) $ is the 4-dimensional coupling constant.

On the other hand, the study of topological quantities
( monopole, instanton, \ldots ) may be important
in order to understand the non-perturbative phenomena.
Many people believe that the dual Meissner effect due to
condensation of color magnetic monopoles
is the color confinement mechanism in QCD \cite{thooft1,mandel}.
Polyakov showed that the confinement features
of 3-dimensional SU(2) gauge theory with Higgs fields in a Higgs phase
is explained by monopole gas ( 3-dimensional instanton )
analytically \cite{poly}.

Recently, we studied the contribution of the monopole to the string
tension in abelian projected SU(2) QCD in maximally abelian gauge
and found the following results \cite{eji}.
\begin{enumerate}
\item Both the physical and the spatial string tension can almost be
explained by monopoles alone.
Although the physical string tension vanishes at the critical
temperature, the spatial string tension remain non-vanishing also
in the case of the string tension calculated by the monopole currents.
\item There exist a long monopole loop and some short loops
in the confinement phase.
The long monopole loop alone is important for the physical string tension.
\item The physical string tension and the long monopole loop
disappear at the same temperature ( $T_{c}$ ).
In the high temperature phase, there are short monopole loops only
and the total monopole number is small.
\end{enumerate}
These results suggest that the important monopole loops are different
in the cases of the spatial string tension and the physical one
and the spatial string tension is produced by small number of monopoles.

The aim of this report is to find what kind of monopole is responsible for
the spatial string tension in the high temperature phase of 4-dimensional QCD
and to study the relation to the Polyakov's monopole gas in the
continuum theory.

\section{Wrapped monopole loop and dimensional reduction}

Polyakov \cite{poly} proved that monopole (instanton) gas can explain
the string tension in 3-dimensional SU(2) gauge theory with Higgs fields.
Hence it is expected that monopole gas may be important for the spatial
string tension also in the dimensional reduced 3-dimensional
effective theory.
When we consider dimensional reduction in 4-dimensional SU(2) gauge theory,
static monopole loops correspond to the monopole gas
in the 3-dimensional effective theory.
The static monopole loops are closed by periodic boundary in the time
direction.
We call these loops wrapped monopole loops.

Because the temperature is the inverse of the lattice size
in the time direction,
the monopole currents may become easier to wrap by the periodic boundary
as the temperature becomes higher.
It may be able to explain the scaling behavior of the spatial string tension.
The spatial string tension becomes larger as the temperature rises.
Since the only difference between time and space directions on the lattice
is the lattice size,
the asymmetry of the physical and spatial string tension must be reduced to
the difference of the boundaries.

In order to know if the spatial string tension at high temperature is
given by the monopole gas in the 3-dimensional theory,
we have checked two points:
\begin{enumerate}
\item
Do only the wrapped monopole loops produce the spatial string tension?
\item
How is the temperature dependence of the wrapped monopole current density?
Does this value remain non-vanishing in the continuum limit?
\end{enumerate}

Our method is the following.
We perform usual Monte-Carlo simulations of SU(2) gauge theory
using the Wilson action.
An abelian theory is extracted by abelian projection
\cite{thooft2,kron}.
The partial gauge fixing is done in the maximally abelian gauge
in which the quantity
\begin{eqnarray}
R=\sum_{s, \mu}{\rm Tr}\Big(\sigma_3 U(s, \mu)
              \sigma_3 U^{\dagger}(s, \mu)\Big)
\end{eqnarray}
is maximized as much as possible by gauge transformation \cite{kron}.
An abelian link field $u(s, \mu)$ is decomposed
from the gauge fixed SU(2) link variable $U(s, \mu)$ as follows:
\begin{eqnarray}
U(s, \mu) = c(s,\mu)u(s,\mu),
\end{eqnarray}
\begin{eqnarray}
c(s, \mu) =  \left( \begin{array}{cc}
             \sqrt{1-|c_{\mu}(s)|^{2}} & -c^{\ast}_{\mu}(s) \\
             c_{\mu} (s) & \sqrt{1-|c_{\mu}(s)|^{2}}
             \end{array} \right),
\hspace{1cm}
u(s, \mu) =  \left( \begin{array}{cc}
             e^{i \theta_{\mu}(s)} & 0 \\
             0 & e^{-i \theta_{\mu}(s)}
             \end{array} \right),
\end{eqnarray}
where $ \theta_{\mu}(s) $ is the abelian gauge field.
$c_{\mu} (s)$ is a complex matter field.
The monopole current $ k_{ \mu }(s) $ is defined as
\begin{eqnarray}
k_{\mu}(s) &=&
(1/4\pi)\epsilon_{\mu\alpha\beta\gamma}\partial_{\alpha}
\bar{\Theta}_{\beta\gamma}(s) \\
\Theta_{ \mu \nu }(s) &=& \partial_{\mu} \theta_{\nu} (s)
- \partial_{\nu} \theta_{\mu} (s) \nonumber \\
 &=& \bar{\Theta}_{ \mu \nu }(s) + 2 \pi n_{\mu \nu}(s)  \nonumber \\
 & & \hspace{-10mm} ( - \pi < \bar{\Theta}_{\mu \nu} \leq \pi ,
\hspace{5mm} n_{\mu \nu} : {\rm integer }) \nonumber
\end{eqnarray}
following Degrand-Toussaint \cite{degrand}.

We define wrapping number of every cluster of connected
monopole currents as follows:
\begin{eqnarray}
{\rm ( wrapping \, number ) }
= \frac{1}{N_{t}} \sum_{\rm \{ cluster \} } k_{4}(s),
\end{eqnarray}
where $ \sum_{\rm \{ cluster \} } $ means summing up in a cluster,
$ N_{t} $ is the lattice size in the time direction.
Non-wrapped monopole loops closed without using periodic boundary condition
give the vanishing wrapping number, since there are the same number of
monopole currents taking $k_{4} = \pm 1$.
On the other hand, if the monopole currents are wrapped
by the periodic boundary, $\sum_{\rm \{ cluster \} } k_{4}(s)$
can take the value of $N_{t}$ times integer.
When a monopole current belongs to a cluster which has non-zero
wrapping number, we regarded it as a wrapped monopole current.

In the confinement phase, the difference between the wrapped monopole
currents and the non-wrapped monopole currents is unclear
since about half of the monopole currents are connected into
one long monopole loop.
However, in the deconfinement phase, there are short loops only
and we can discriminate both monopole currents.

\section{Wrapped monopole contribution to the spatial string tension}

We have studied the wrapped monopole contribution
to the spatial string tension in the high temperature phase.
We have considered the monopole contribution to the Wilson loop
as discussed in \cite{shiba,stack}.

First, we extract abelian component by performing abelian projection
in the maximally abelian gauge.
In this gauge, the spatial string tension can be reproduced by
residual abelian link variables.
Next, we decompose the abelian Wilson loop $W$, which is the Wilson loop
composed of abelian link variables, into two parts $W_{1}$ ( photon part )
and $W_{2}$ ( monopole part ) as follows:
\begin{eqnarray}
W\    & = & \exp\{i \sum \theta_{\mu}(s) J_{\mu}(s)\} \\
      & = & W_{1} \cdot W_{2} \label{w12}\\
W_{1} & = & \exp\{-i \sum \partial'_{\mu}\bar{\Theta}_{\mu\nu}(s)
D(s-s')J_{\nu}(s')\} \label{wph} \\
W_{2} & = & \exp\{2\pi i \! \sum \! k_{\beta}(s)D(s \! - \!s' \!)\frac{1}{2}
\epsilon_{\alpha\beta\rho\sigma}\partial_{\alpha}M_{\rho\sigma}( \! s'\! )\},
\label{wmon}
\end{eqnarray}
where $D(s-s')$ is the lattice Coulomb propagator.
$J_{\mu}(s)$ is an external current corresponding to the Wilson loop and
$M_{\mu \nu}(s)$ is an antisymmetric variable taking $ \pm 1 $
on a surface with the Wilson loop boundary as
$J_{\nu}(s) = \partial'_{\mu} M_{\mu\nu}(s)$.
The string tension is obtained from these Wilson loops
by the least square fit.
We have assumed that the static quark anti-quark potential is given
by linear + Coulomb + constant terms.
The Creutz ratio of the size $R \times S$ on the lattice is
\begin{eqnarray}
\chi (R,S) = \chi_{0} + \chi_{1} ( \frac{1}{R(R-1)} + \frac{1}{S(S-1)} )
+ \chi_{2} ( \frac{1}{R(R-1)S(S-1)} ),
\end{eqnarray}
where $\chi_{0}$ is the spatial string tension in the lattice unit.
We have adopted this fitting function.
The spatial string tension $\sigma_{s}(T)$ can be expressed in units of the
critical temperature ($T_{c}$)
using the relation between the lattice spacing and
the temperature \cite{bali}:
\begin{eqnarray}
\frac{\sqrt{\sigma_{s} (T)}}{T_{c}} = \sqrt{\chi_{0} (T)} N_{tc},
\end{eqnarray}
where $N_{tc}$ is the critical lattice size at each $\beta$.

Notice that we can see that a time-like monopole current $(k_{4})$ such as
the static monopole current does not affect the physical string tension
and it contributes only to the spatial one as seen from equation (\ref{wmon}).
When we measure the physical string tension, we use the Wilson loop having
the antisymmetric variable $M_{i 4}$ or $M_{4 i}$, where $i = 1,2,3$.
Because of the antisymmetric $\epsilon$ tensor,
the physical string tension is evaluated by
the space-like monopole currents only.

The monopole contribution in the deconfinement phase is a
little lower than the full one, but it almost reproduce the behavior of
the full one in the maximally abelian gauge \cite{eji}.

\begin{figure}[htb]
\epsfxsize=\textwidth
\begin{center}
\leavevmode
\epsfbox{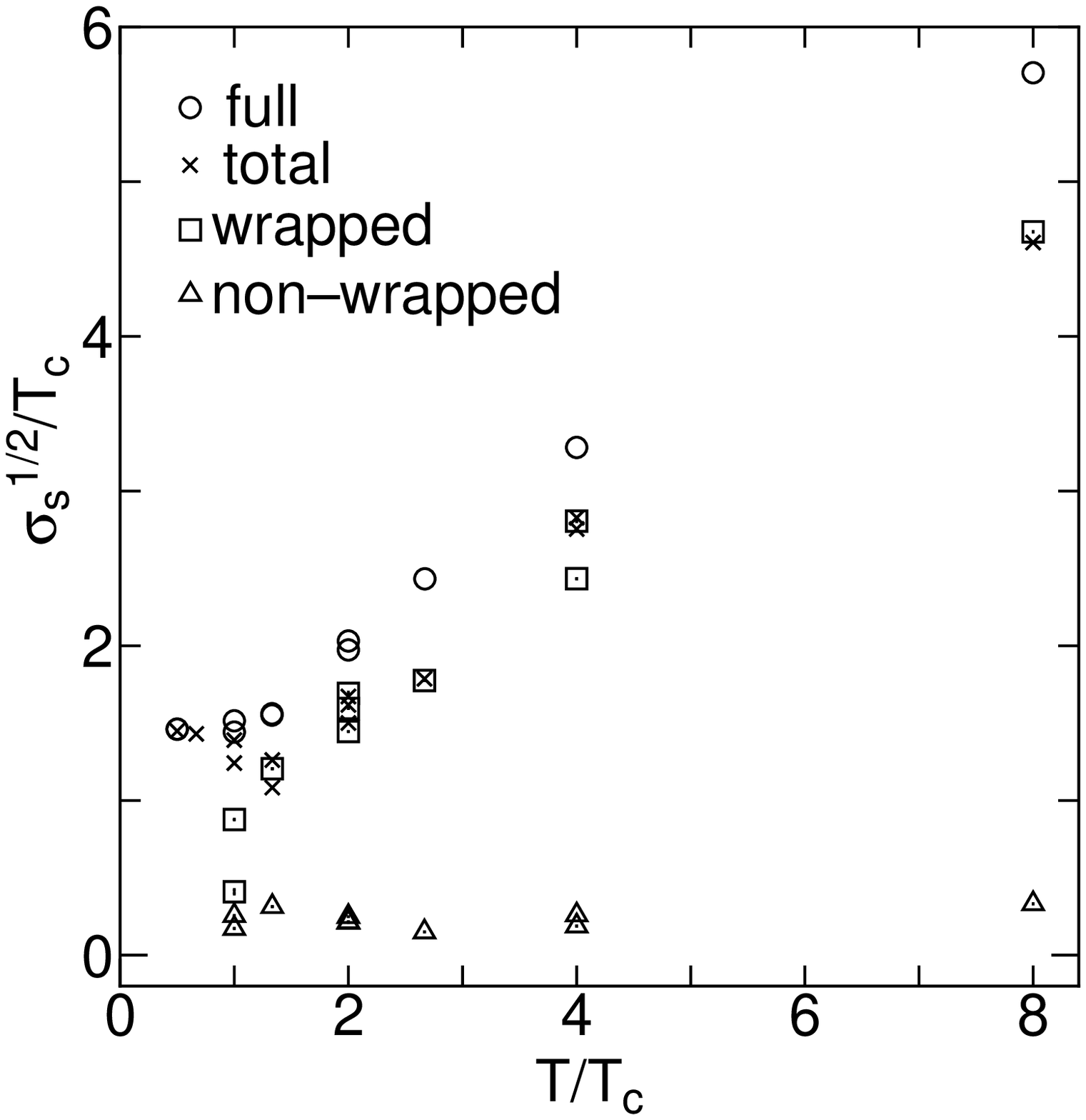}
\end{center}
\caption{
The full spatial string tension (circle),
the total monopole contribution (cross),
the wrapped monopole contribution (square) and
the non-wrapped monopole contribution (triangle).
The full one is cited from [2].}
\label{spst}
\end{figure}

Here, we calculate the wrapped monopole contribution
and the non-wrapped monopole contribution
to the spatial string tension separately
in the high temperature phase.
We have performed Monte-Carlo simulations on $ 24^{3} \times N_{t} $ lattices
with the periodic boundary condition,
$ N_{t} = \{ 2,4,6,8 \} $,
at $ \beta = \{ 2.30, 2.51, 2.74 \} $
which are the critical $\beta$ for $ N_{t} = \{ 4, 8, 16 \} $ respectively.
Measurements have been done every 50 sweeps
after a thermalization of 2000 sweeps.
We have taken 50 configurations for measurements.
The data are plotted in Fig.\ \ref{spst}.
These data show that
the spatial string tension from the wrapped monopole is almost the same
as that from total monopole loops and show that
the non-wrapped loops do not contribute to the spatial string tension.

\section{Scaling behavior of wrapped monopole density}

The spatial string tension looks non-vanishing in the continuum limit.
If the wrapped monopole produces the spatial string tension,
the wrapped monopole density must remain
non-vanishing in the continuum limit.
We have investigated the monopole density on a lattice of the size
$ N_{s}^{3} \times N_{t} $.
The monopole density $ \rho (T) $
is defined in the physical unit as follows:
\begin{eqnarray}
\rho (T) = \frac{ \sum |k_{\mu}(s)|a }{ (N_{s}a)^{3}(N_{t}a) }.
\end{eqnarray}
Here $a$ is the lattice spacing.
Considering the relation between the temperature and the lattice size
$ N_{t} a = 1 / T $, we can rewrite the monopole density as follows:
\begin{eqnarray}
\rho (T) = \frac{ \sum |k_{\mu}(s)| N_{tc}^{3} }{ N_{s}^{3} N_{t} } T_{c}^{3},
\end{eqnarray}
where $T_{c}$ is the critical temperature and
$N_{tc}$ is the critical lattice size at each $\beta$.

We have measured the temperature dependence and the $\beta$ dependence
both of the total and the wrapped monopole densities
varying both $\beta$ and $N_{t}$
on $24^{3} \times N_{t}$ lattices. \mbox{($\beta = \{ 2.30, 2.51, 2.74 \}$,
$ N_{t} = \{ 2, 4, 6, 8, 12 \} $)}

\begin{figure}[htb]
\epsfxsize=\textwidth
\begin{center}
\leavevmode
\epsfbox{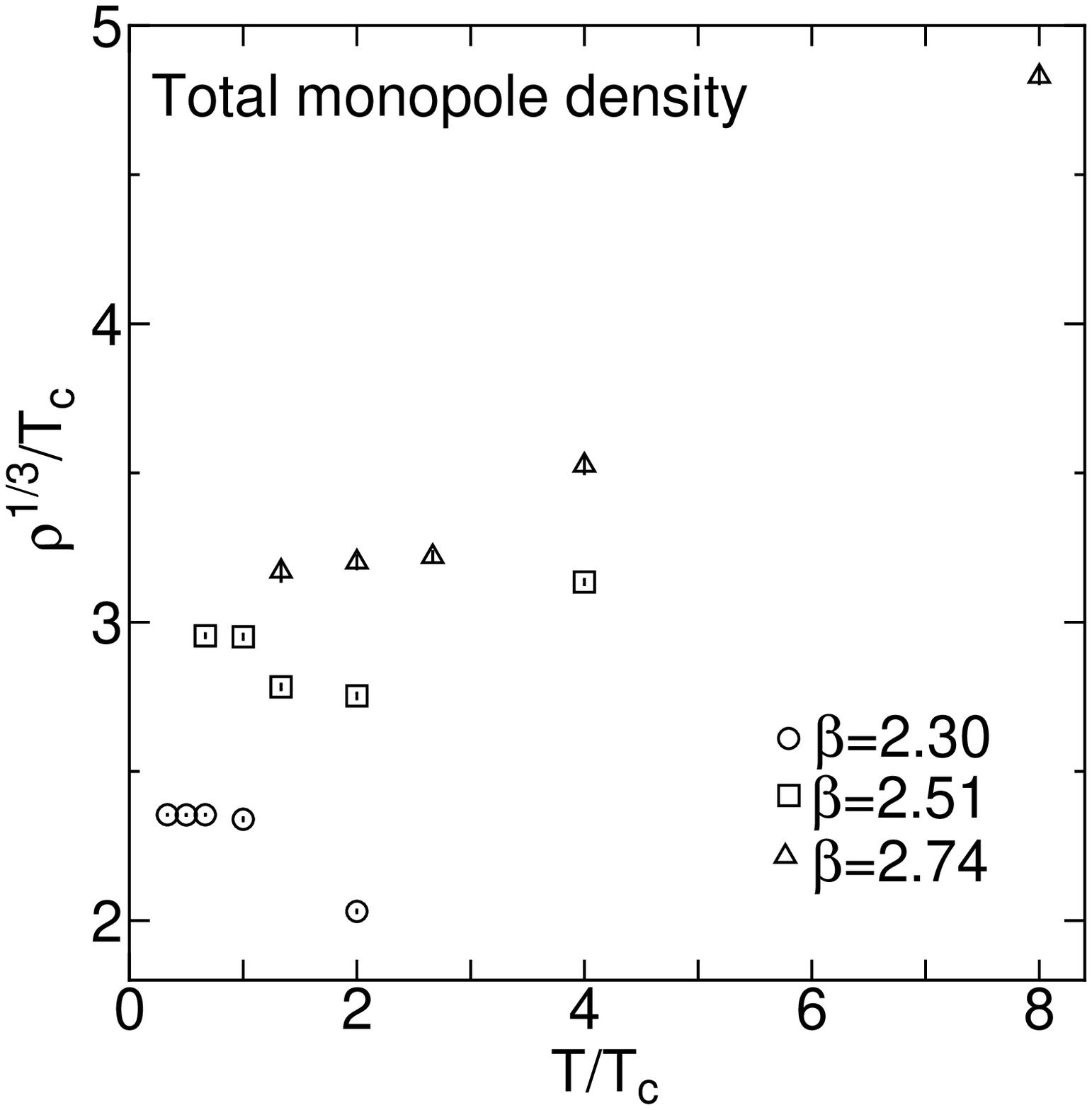}
\end{center}
\caption{The temperature dependence of the total monopole density
at $ \beta = 2.30 $ (circle), $ \beta = 2.51 $ (square) and
$ \beta = 2.74 $ (triangle)
on $ N_{t} = 2, 4, 6, 8$ and $12$ lattices respectively.}
\label{mpd}
\end{figure}

\begin{figure}[htb]
\epsfxsize=\textwidth
\begin{center}
\leavevmode
\epsfbox{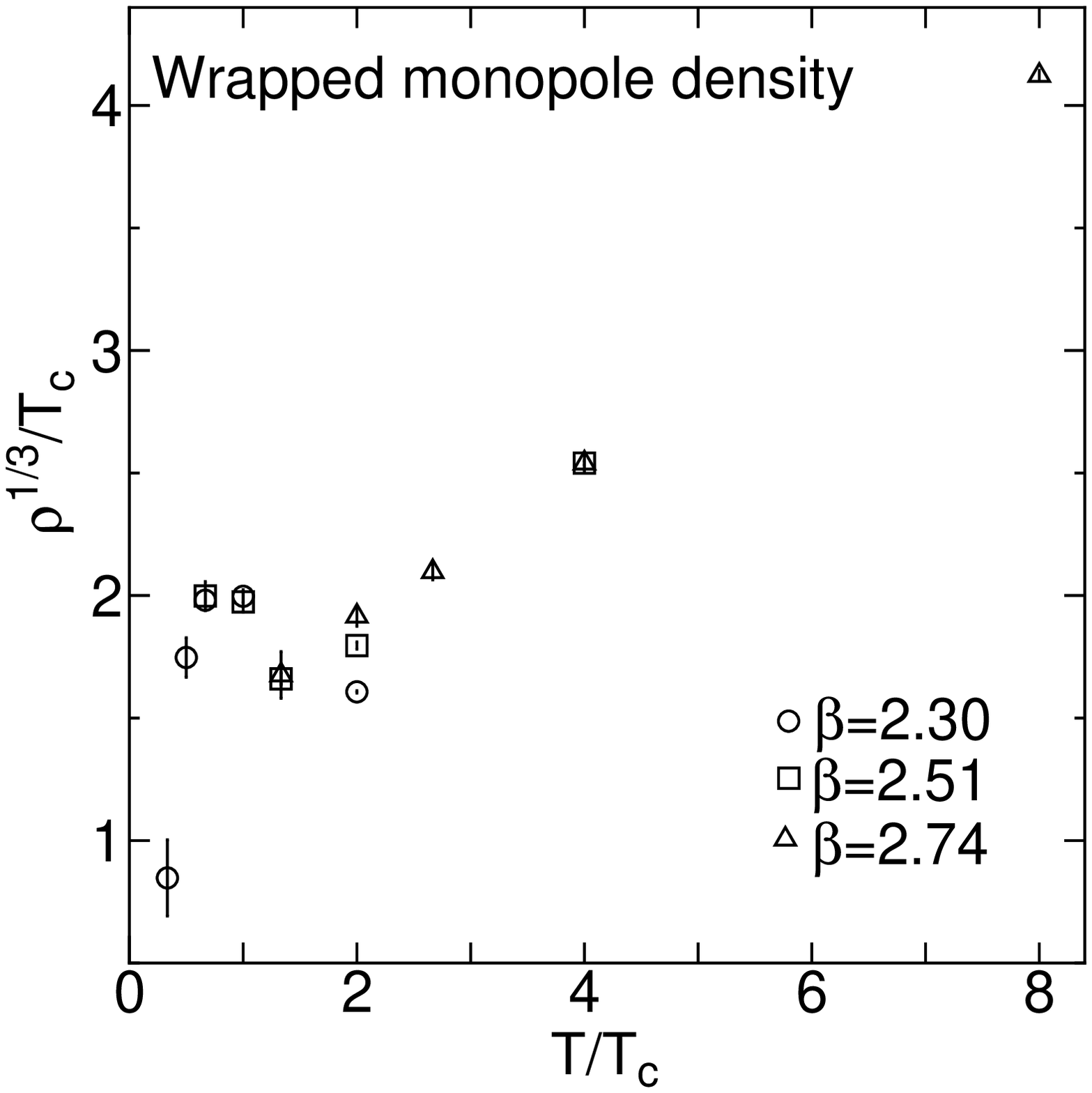}
\end{center}
\caption{The temperature dependence of the wrapped monopole density
at $ \beta = 2.30 $ (circle), $ \beta = 2.51 $ (square) and
$ \beta = 2.74 $ (triangle)
on $ N_{t} = 2, 4, 6, 8$ and $12$ lattices respectively.}
\label{wmd}
\end{figure}

The data of the total and the wrapped monopole densities are
shown in Fig.\ \ref{mpd} and in Fig.\ \ref{wmd}.
The total monopole density does not show good scaling behavior.
It depends not only on $T$ but also on $\beta$.
On the other hand, the wrapped monopole density in the high temperature phase
is independent of $\beta$,
and looks remaining in the continuum limit.
The wrapped monopole density seems to be proportional to $T^{3}$
at high temperature.
This scaling behavior is similar to that of the spatial string tension.

\section{ Conclusions and Discussion }

We have found the following results
by the numerical studies of Monte-Carlo simulations.
The spatial string tension can almost be reproduced by the
wrapped monopole loops closed by the periodic boundary
in the time direction in the high temperature phase.
The wrapped monopole density is independent of $\beta$ and
appears non-vanishing
in the continuum limit. Moreover the scaling behavior of
the wrapped monopole density is similar to that of the spatial string tension.
These results suggest that the spatial string tension at high temperature
is produced by the monopole gas in the effective 3-dimensional theory.

In Polyakov's analytical calculation,
it is essential that the Higgs field has non-zero vacuum expectation value.
On the other hand,
the study of the scaling behavior of the spatial string tension
\cite{bali,karsch} suggested that
the Higgs-sector in dimensionally reduced QCD
does not contribute significantly to the spatial string tension.
Furthermore, It was reported that the Higgs field does not have
a non-zero vacuum expectation value \cite{kark}.
We expect that the monopole gas can be discussed without
the expectation value of the Higgs field
as we are discussing originally the monopole currents in 4-dimensional
QCD without a Higgs field.
If one wants to study the confinement mechanism using the classical solution
of the monopole gas, the non-zero expectation value of the Higgs field is
necessary. However, in the scheme of the abelian projection,
the monopoles are produced dynamically and such monopoles are expected to
play an important role for the confinement mechanism.
Our results suggest that the
spatial string tension has a close relation to the monopole gas.

\section*{ Acknowledgments }

The author thanks T.Suzuki, Y.Matsubara and S.Kitahara for fruitful
discussions and comments.
Prof. T.Suzuki suggested that the spatial string tension at high temperature
has a relation to the Polyakov's monopole gas.
The calculations were performed on Fujitsu VPP500 at
the institute of Physical and Chemical Research (RIKEN) and
National Laboratory for High Energy Physics (KEK).

\end{document}